\documentstyle[prd,aps]{revtex}

\input epsf
\epsfverbosetrue

\def\Journal#1#2#3#4{{#1} {\bf #2} (#4) #3}
\def\MPL{Mod. Phys. Lett. A}

\def\NPB{Nucl. Phys. B}

\def\NPSUPPL{Nucl. Phys. Proc. Suppl.}
\def\PLB{{Phys. Lett.} B}

\def\PLBOLD{Phys. Lett.}
\def\PRL{Phys. Rev. Lett.}
\def\RMP{Rev. Mod. Phys.}
\def\PRD{Phys. Rev. D}

\def\PTP{Prog. Theor. Phys.}
\def\JHEP{JHEP}

\def\JETPUSSR{JETP (USSR)}
\def\ZETP{Zh. Eksp. Teor. Piz.}
\def\TNYAS{Trans. New York Acad. Sci.}

\def\mapgeq{\mathbin{\lower.3ex\hbox{$\buildrel>\over{\smash{\scriptstyle\sim}\vphantom{_x}}$}}}
\def\mapleq{\mathbin{\lower.3ex\hbox{$\buildrel<\over{\smash{\scriptstyle\sim}\vphantom{_x}}$}}}
\def\mapgeqeq{\mathbi{\lower.3ex\hbox{$\buildrel>\over{\smash{\scriptstyle\approx}\vphantom{_2}}$}}}
\def\mapleqeq{\mathbin{\lower.3ex\hbox{$\buildrel<\over{\smash{\scriptstyle\approx}\vphantom{_2}}$}}}

 \mathchardef\#="0023
 \mathchardef\$="0024
 \mathchardef\%="0025
 \mathchardef\ddash="705C
 
 \mathchardef\lwavy="336E
 \mathchardef\rwavy="336F
 \mathchardef\biglwavy="331A
 \mathchardef\bigrwavy="331B
 \mathchardef\bigglwavy="3328
 \mathchardef\biggrwavy="3329
 \mathchardef\littlesum="0350

\tighten
\draft
\begin{document} 
\bibliographystyle{prsty}

\title{
Large Solar Neutrino Mixing in an Extended Zee Model\footnote{to be published in Int. J. Mod. Phys. A (2002)}
}

\author{
Teruyuki Kitabayashi$^a$
\footnote{E-mail:teruyuki@post.kek.jp}
and Masaki Yasu${\grave {\rm e}}^b$
\footnote{E-mail:yasue@keyaki.cc.u-tokai.ac.jp}
}

\address{\vspace{5mm}$^a$
{\sl Accelerator Engineering Center} \\
{\sl Mitsubishi Electric System \& Service Engineering Co.Ltd.} \\
{\sl 2-8-8 Umezono, Tsukuba, Ibaraki 305-0045, Japan}
}
\address{\vspace{2mm}$^b$
{\sl Department of Natural Science\\School of Marine
Science and Technology, Tokai University}\\
{\sl 3-20-1 Orido, Shimizu, Shizuoka 424-8610, Japan\\and\\}
{\sl Department of Physics, Tokai University} \\
{\sl 1117 KitaKaname, Hiratsuka, Kanagawa 259-1292, Japan}}
\date{TOKAI-HEP/TH-0107, December, 2001}
\maketitle

\begin{abstract}
The Zee model, which employs the standard Higgs scalar ($\phi$) with its duplicate ($\phi^\prime$) and a singly charged scalar ($h^+$), can utilize two global symmetries associated with the conservation of the numbers of $\phi$ and $\phi^\prime$, $N_{\phi,\phi^\prime}$, where $N_\phi+N_{\phi^\prime}$ coincides with the hypercharge while $N_\phi-N_{\phi^\prime}$ ($\equiv X$) is a new conserved charge, which is identical to $L_e-L_\mu-L_\tau$ for the left-handed leptons.  Charged leptons turn out to have $e$-$\mu$ and $e$-$\tau$ mixing masses, which are found to be crucial for the large solar neutrino mixing.  In an extended version of the Zee model with an extra triplet Higgs scalar (s), neutrino oscillations are described by three steps: 1) the maximal atmospheric mixing is induced by democratic mass terms supplied by $s$ with $X$=2 that can initiate the type II seesaw mechanism for the smallness of these masses; 2) the maximal solar neutrino mixing is triggered by the creation of radiative masses by $h^+$ with $X$ = 0; 3) the large solar neutrino mixing is finally induced by a $\nu_\mu$-$\nu_\tau$ mixing arising from the rotation of the radiative mass terms as a result of the diagonalization that converts $e$-$\mu$ and $e$-$\tau$ mixing masses into the electron mass.
\end{abstract}
\vspace{4mm}

The experimental confirmation of neutrino oscillations has been first given by the Super-Kamiokande collaboration (Super-K) \cite{Kamiokande} for atmospheric neutrinos and the clear evidence of the solar neutrino oscillations has lately been  released by the SNO collaboration \cite{SNO} in their observation of a non-electron flavor active neutrino component in the solar flux.  Atmospheric neutrinos are known to exhibit maximal mixing of $\nu_\mu$ with $\nu_\tau$ and it has also been suggested for solar neutrinos that solutions with large mixing angles are favored while solutions with small mixing angles are disfavored in the recent Super-K report on the detailed analysis of solar neutrino oscillations \cite{RecentSK}. Therefore, both observed oscillations are characterized by large neutrino mixings with $\sin^22\theta_{23} \sim$ 1.0 and $\Delta m^2_{atm} \sim 3.0 \times 10^{-3}$ eV$^2$ \cite{RecentAtmAnalysis} for atmospheric neutrinos and with $\sin^22\theta_{12} \sim$ 0.8 and $\Delta m^2_\odot \sim 4.5 \times 10^{-5}$ eV$^2$ \cite{RecentSolarAnalysis} for solar neutrinos corresponding to the most favorable large mixing angle (LMA) MSW solution.

Neutrino oscillations occur if neutrinos have masses \cite{EarlyMassive}.  Since the oscillation data imply very tiny masses of ${\mathcal{O}}$(10$^{-2}$) eV, one has to explain why neutrinos have such tiny masses \cite{Seesaw,Zee}.  Neutrinos can be massive Majorana particles without their right-handed partners, whose masses are radiatively generated by the mechanism as discussed by Zee \cite{Zee}.  This radiative mechanism indeed characterizes an underlying physics generating tiny neutrino masses. To be phenomenologically consistent, it is suggested \cite{Useful} to use the bimaximal mixing scheme \cite{BimaximalMixing,NearlyBiMaximal} based on a global $L_e-L_\mu-L_\tau$ ($\equiv L^\prime$) symmetry \cite{EarlierLprime,Lprime}. Since the Zee model only supplies flavor-off-diagonal mass terms, its bimaximal structure is inevitable to accommodate observed neutrino oscillations.  As a result, we end up with the maximal solar neutrino mixings with $\sin^22\theta_{12} \sim 1$ \cite{ZeeMaximal}. Recent extensive analyses on solar neutrino data have, however, indicated that such a maximal solar mixing angle is not well compatible with the data, which prefer $\sin^22\theta_{12} \sim 0.8$ \cite{RecentSolarAnalysis}.

In this paper, we extend the Zee model with the $L^\prime$ symmetry to accommodate the LMA solution without the maximal solar neutrino mixing \cite{ZeeLMA,ZeeLMA2}.  The original Zee model consists of leptons ($\psi^{e,\mu,\tau}_L$ and $\ell^{e,\mu,\tau}_R$), two $SU(2)_L$-doublet Higgs scalars ($\phi$, $\phi^\prime$) and one $SU(2)_L$-singlet charged scalar ($h^+$), where $\phi^\prime$ is constrained to couple to no leptons.  However, such a constraint on the interactions of $\phi^\prime$ should be determined by a certain underlying symmetry.  We introduce the invariance imposed by two global symmetries associated with the numbers of $\phi$ and $\phi^\prime$, $N_\phi$ and $N_{\phi^\prime}$.  It will be shown that $N_\phi+N_{\phi^\prime}$ coincides with the hypercharge.  The orthogonal combination, $N_\phi-N_{\phi^\prime}$, turns out to be a new conserved quantum number, which is identical to $L_e-L_\mu-L_\tau$ for the left-handed leptons. As a result, $\phi^\prime$ is allowed to couple to $\mu$ and $\tau$ and, at the same time, $\phi$ is constrained to have couplings of $e$-$\mu$ and $e$-$\tau$ \cite{ZeeLMA}.  Since off-diagonal mass terms for charged leptons are generated by $\phi$, neutrino mixings are affected by the diagonalization of these charged lepton mass terms, which will be used to explain $\sin^22\theta_{12}\neq 1$. 

In order to realize $\sin^22\theta_{12} \neq 1$, the simplest resolution is to include flavor-diagonal mass terms.  It is because the main source of $\sin^22\theta_{12} \approx 1$ in the Zee model comes from the constraint on neutrino masses $m_{1,2,3}$ of $m_1+m_2+m_3$=0 specific to flavor-off-diagonal mass terms.  We employ an additional $SU(2)_L$-triplet Higgs scalar ($s$):\cite{Triplet}
\begin{eqnarray}\label{Eq:s-Higgs}
&s=\left( \begin{array}{cc}
  s^+ &  s^{++}\\
  s^0 &  -s^+
\end{array} \right)
\end{eqnarray}
to supply flavor-diagonal mass terms.  Tree level neutrino masses are generated by the vacuum expectation value of $s$, $\langle 0 \vert s^0 \vert 0 \rangle$, via interactions of ${\overline {\psi^c_L}} s \psi_L$, where the subscript $c$ denotes the charge conjugation including the $G$-parity of $SU(2)_L$. The smallness of the neutrino masses is ascribed to that of $\langle 0 \vert s^0 \vert 0 \rangle$, which is given by $\sim \mu (\langle 0 \vert \phi \vert 0 \rangle/m_s)^2$ with $\langle 0 \vert \phi \vert 0 \rangle$ produced by the combined effects of $\mu\phi^\dagger s \phi^c$ and $m^2_s$Tr$(s^\dagger s)$, where $\mu$ and $m_s$ are mass parameters. The type II seesaw mechanism \cite{Type2SeeSaw} can be used to ensure tiny neutrino masses by the dynamical requirement of $\vert \langle 0 \vert \phi \vert 0 \rangle \vert \ll m_s$ with $\mu\sim m_s$.

In the next section, we present a possible neutrino mass matrix to be diagonalized by two mixing angles, $\theta_{12}$ for the $\nu_e$-$\nu_\mu$ mixing and $\theta_{23}$ for the $\nu_mu$-$\nu_\tau$ mixing, from which one can find which elements make $\sin^22\theta_{12}$ less than unity.  In Sec.III, we explain the role of two conserved charges of $N_{\phi,\phi^\prime}$ and show the Yukawa and Higgs interactions, which provide radiative neutrino masses for $\nu_e$-$\nu_\mu$ and $\nu_e$-$\nu_\tau$ and charged lepton masses for $e$-$\mu$ and $e$-$\tau$.  The diagonalization of the charged lepton masses is performed in Sec.IV and also shown is sufficient suppression of possible flavor-changing interactions, which are generated by $\phi$ and $\phi^\prime$ since charged leptons simultaneously couple to these Higgs scalars. Solar neutrino oscillations are found to exhibit the large mixing of $\sin^22\theta_{12}\sim 0.8$ as a result of the rotation.  The last section is devoted to summary and discussions.

\section{Neutrino Mass Texture}
\noindent
Before discussing how neutrino masses and oscillations really come out, we first examine which flavor neutrino mass terms affects the deviation of $\sin^22\theta_{12}$ from unity.  The mass matrix given by $M^\nu$ of the form of \cite{ZeeLMA,MassMatrix}
\begin{eqnarray}\label{Eq:M_NU}
&M^\nu = \left( \begin{array}{ccc}
  a           &  b     & c (= -t_{23}b)\\
  b           &  d     & e\\
  c      &  e     & f(=d+\left( t^{-1}_{23}-t_{23}\right) e)
\end{array} \right)
\end{eqnarray}
can be diagonalized by $U_{MNS}$ defined by
\begin{eqnarray}\label{Eq:U_MNS}
&U_{MNS}=\left( \begin{array}{ccc}
  \cos\theta_{12} &  \sin\theta_{12}&   0\\
  -\cos\theta_{23}\sin\theta_{12}&  \cos\theta_{23}\cos\theta_{12}&  \sin\theta_{23}\\
  \sin\theta_{23}\sin\theta_{12}&  -\sin\theta_{23}\cos\theta_{12} & \cos\theta_{23}\\
\end{array} \right),
\end{eqnarray}
where $t_{23}$ = $\sin\theta_{23}/\cos\theta_{23}$, which transforms $\vert \nu_{mass}\rangle$ = ($\nu_1, \nu_2, \nu_3$)$^T$ with masses of ($m_1, m_2, m_3$) into $\vert \nu_{weak}\rangle$ = ($\nu_e, \nu_\mu, \nu_\tau$)$^T$: $\vert \nu_{weak}\rangle = U_{MNS} \vert \nu_{mass}\rangle$.  The masses and $\theta_{12}$ are calculated to be \cite{NewWork}:
\begin{eqnarray}
&&
m_1 = a-\frac{1}{2}\sqrt{\frac{b^2+c^2}{2}}\left( x+\eta\sqrt{x^2 + 8}\right), \quad m_2 = \left( \eta \rightarrow -\eta ~{\rm in}~m_1\right), 
\nonumber \\
&&
m_3 = d + t^{-2}_{23}\left( d-a +x\sqrt{\frac{b^2+c^2}{2}}\right),
\label{Eq:NuMasses123} \\
&&
\sin^22\theta_{12}=\frac{8}{8+x^2}~{\rm with}~x=\frac{a-d+t_{23}e}{\sqrt{(b^2+c^2)/2}},
\label{Eq:Angle12}
\end{eqnarray}
where $\vert m_1\vert < \vert m_2\vert$ is always maintained by adjusting the sign of $\eta$ (= $\pm 1$). The relation of Eq.(\ref{Eq:Angle12}) shows that the significant deviation of $\sin^22\theta_{12}$ from unity is only possible if $\vert x \vert = {\cal O}(1)$, namely, $( a-d+t_{23}e)^2 = {\cal O}(b^2+c^2)$. In our subsequent discussions, the solution of $a=0$ and $(d-t_{23}e)^2 = {\cal O}(b^2+c^2)$ is realized.  

Let us briefly examine Eqs.(\ref{Eq:NuMasses123}) and (\ref{Eq:Angle12}) in the Zee model.  Since the Zee mass matrix is parameterized by $a=d=0$, leading to $m_1+m_2+m_3$ = 0, we find that 
\begin{eqnarray}\label{Eq:MassSquared}
&&
\Delta m^2_{atm} = m^2_3-m^2_2 
=\left[ x^2-\left( 4+\vert x \vert\sqrt{x^2+8}\right) \right]\frac{b^2+c^2}{4},
\nonumber \\
&&
\Delta m^2_\odot = m^2_2-m^2_1
=\vert x \vert \sqrt{x^2+8}\frac{b^2+c^2}{2}.
\end{eqnarray}
where $\vert t_{23}\vert=1$ is used for $\Delta m^2_{atm}$.  It is obvious that the requirement of $\vert x \vert$ = ${\cal O}(1)$ for the large solar neutrino mixing gives $\Delta m^2_\odot = {\cal O}(\Delta m^2_{atm})$, which contradicts with the observed result.  More explicitly, the mixing angle $\theta_{12}$ is computed to be:\cite{h_phenom1}
\begin{eqnarray}\label{Eq:Angle_NU}
&&
\sin^2 2\theta_{12} = \frac{4 \vert m_1m_2\vert}{(\vert m_1\vert + \vert m_2\vert)^2},
\end{eqnarray}
Since $(m_2-m_1)^2 \geq m^2_3$ is derived, $\sin^2 2\theta_{12} \approx 1$ is expected because of $\vert\Delta m^2_{atm} \vert \gg \vert\Delta m^2_\odot \vert$ implying $\vert m_1\vert \approx \vert m_2\vert$.  In fact,  one can readily find that
\begin{eqnarray}\label{Eq:M_NU_MASS}
&&
m^2_1 = \left( 2M - \Delta m^2_{atm} -2 \Delta m^2_\odot\right)/3,
\quad
m^2_2 = \left( 2M - \Delta m^2_{atm} + \Delta m^2_\odot\right)/3,
\nonumber \\
&&
m^2_3 = \left( 2M + 2\Delta m^2_{atm} + \Delta m^2_\odot\right)/3,
\end{eqnarray}
where $M$ = $\sqrt{(\Delta m^2_{atm})^2+\Delta m^2_{atm}\Delta m^2_\odot+(\Delta m^2_\odot)^2}$ reduced to $(m^2_1+m^2_2+m^2_3)/2$ as expected, which can be derived by the use of $m_1+m_2+m_3=0$.  The approximation of $\vert\Delta m^2_{atm}\vert \gg\vert \Delta m^2_\odot\vert$, finally, gives $\sin^22\theta_{12}$ expressed as:\cite{ZeeMaximal}
\begin{eqnarray}\label{Eq:Sin12ZEE}
&&
\sin^22\theta_{12} \approx 1-\frac{1}{16}\left( \frac{\Delta m^2_\odot}{\Delta m^2_{atm}}\right)^2.
\end{eqnarray}
This result is entirely based on the constraint of $m_1+m_2+m_3=0$, which enables us to relate $m^2_{1,2,3}$ to $\Delta m^2_{atm}$ and $\Delta m^2_\odot$,  This  constraint is violated by the inclusion of diagonal masses supplied by $s$.

We start with the ``ideal" solution \cite{Moha} with $t_{23}=\pm 1$ ($\equiv\sigma$) given by
\begin{eqnarray}\label{Eq:SimplestNuMass1}
&M^\nu_{ideal}=
\left( \begin{array}{ccc}
  0            &  0            & 0\\
  0            &  d            & \sigma d\\
  0    &  \sigma d     & d
\end{array} \right),
\end{eqnarray}
which provides $m_1=m_2 =0$ and $m_3 = 2d$ and the maximal atmospheric mixing.  The deviation from this solution that yields $b(=b_{rad})\neq 0$ and $c(=c_{rad})\neq 0$ as $\nu_e\nu_\mu$- and $\nu_e\nu_\tau$-terms is caused by radiative effects, leading to $M^\nu_1$:
\begin{eqnarray}\label{Eq:SimplestNuMass12}
&M^\nu_1=
\left( \begin{array}{ccc}
  0            &  b_{rad}           & c_{rad}\\
  b_{rad}            &  d            & \sigma d\\
  c_{rad}    &  \sigma d     & d
\end{array} \right),
\end{eqnarray}
which provides the bimaximal neutrino mixing for $b_{rad} = -\sigma c_{rad}$. The diagonalization of the charged lepton masses generates an extra $\nu_\mu\nu_\tau$-term (=$d_{rot}$) through the rotation of $b_{rad}$ and $c_{rad}$, which gives $(d-t_{23}e)^2 = {\cal O}(b^2+c^2)$, leading to $M^\nu_2$:
\begin{eqnarray}\label{Eq:SimplestNuMass13}
&M^\nu_2=
\left( \begin{array}{ccc}
  0            &  b_{rad}           & c_{rad}\\
  b_{rad}            &  d            & \sigma d + d_{rot}\\
  c_{rad}    &  \sigma d + d_{rot}   & d
\end{array} \right),
\end{eqnarray}
which finally provides the large solar neutrino mixing.

\section{Model}
To realize interactions that yield the ``ideal" solution of Eq.(\ref{Eq:SimplestNuMass1}) and radiative neutrino masses in the specific entries of $\nu_e\nu_\mu$ and $\nu_e\nu_\tau$ as in Eq.(\ref{Eq:SimplestNuMass12}), we require all interactions to be invariant under the transformations of two symmetries, $U(1)_\phi$ and $U(1)_{\phi^\prime}$ associated with $N_\phi$ and $N_{\phi^\prime}$, where ($N_\phi$, $N_{\phi^\prime}$) = (1, 0) for $\phi$; = (0, 1) for $\phi^\prime$; = (0, $-1$) for $\psi^e_L$; = ($-1$, 0) for $\psi^{\mu,\tau}_L$; = ($-2$, 0) for $\ell^e_R$; = ($-1$, $-1$) for $\ell^{\mu,\tau}_R$; = (1, 1) for $h^+$; and = (2, 0) for $s$. Similarly, the ordinary lepton number, $L$, is given by $L$ = 1 for leptons, $L$ = $-2$ for $h^+$ and $s$ and $L$ = 0 for $\phi$ and $\phi^\prime$.  These quantum numbers are tabulated in TABLE \ref{Tab:QuantumNumber1}.  One can recognize that $N_\phi + N_{\phi^\prime}$ is nothing but the hypercharge ($Y$) and that $N_\phi-N_{\phi^\prime}$ coincides with $L^\prime$ for $\psi^{e,\mu,\tau}_L$.  Because $\phi$ has $N_\phi = Y$ and $N_{\phi^\prime}=0$, this charge assignment is readily extended to quarks such that the conservation of $N_{\phi,\phi^\prime}$ is respected by Yukawa couplings of quarks to $\phi$ if $N_\phi = Y$ and $N_{\phi^\prime} = 0$ are assigned to quarks. But, $\phi^\prime$ is forbidden to couple to quarks. Since leptons couple to both $\phi$ and $\phi^\prime$, our model yields additional contributions to the well-established low-energy phenomenology of leptons including those from dangerous flavor-changing interactions, which cause rare decays such as $\tau\to\mu\mu\mu, \mu e e, \mu\gamma$ and $\mu\to eee, e\gamma$. It will be shown that these interactions are well suppressed.

The Yukawa interactions for leptons are, now, given by
\begin{eqnarray}
-{\cal L}_Y   &=& 
\sum_{i=\mu,\tau}
\left(
f^{\phi}_{ie}{\overline {\psi^i_L}}\phi \ell^e_R +f^{\phi}_{ei}{\overline {\psi^e_L}}\phi \ell^i_R
+
f_{[ei]}{\overline {\left( \psi_L^e \right)^c}}\psi^i_Lh^+
\right)
\nonumber \\
&&
+\sum_{i,j=\mu,\tau}
\left(
f^{\phi^\prime}_{ij} {\overline {\psi^i_L}}\phi^\prime\ell^j_R
+
\frac{1}{2}f_{\{ij\}}{\overline {\left( \psi_L^i \right)^c}}s\psi^j_L
\right)
+ {\rm (h.c.)},
\label{Eq:OurYukawa}
\end{eqnarray}
where $f$'s stand for coupling constants and the subscripts of $[ij]$ and $\{ ij\}$, respectively, denote the symmetrization and antisymmetrization with respect to $i$ and $j$. The masses of $\mu$ and $\tau$ are taken to be diagonal for simplicity.  To meet the ``ideal" solution of Eq.(\ref{Eq:SimplestNuMass1}), we set $f_{\{ \mu\mu\}}$ = $f_{\{ \tau\tau\}}$ = $\sigma f_{\{ \mu\tau\}}$.  Although this solution may require other dynamics or a certain symmetry restriction such as the one from a permutation symmetry of $S_2$ for the $\mu$ and $\tau$ \cite{NewWork,Permutation}, we do not further pursue such appropriate physical reasons.  Instead, we examine how the large solar neutrino mixing is implemented \cite{How} in our radiative mechanism by adopting this ``ideal" solution.

Higgs interactions are described by usual Hermitian terms composed of $\varphi\varphi^\dagger$ ($\varphi$ = $\phi$, $\phi^\prime$, $h^+$, $s$) and by non-Hermitian terms in
\begin{eqnarray}
V_0
& = &
\mu_0\phi^{c\dagger}\phi^\prime h^{+\dagger}
+
\lambda\phi^{c \dagger} s\phi^{\prime c}h^{+\dagger}
+
\mu\phi^\dagger s\phi^c
+
{\rm (h.c.)}, 
\label{Eq:Conserved}
\end{eqnarray}
where $\mu_0$ and $\mu$ represent mass scales and $\lambda$ stands for a Higgs coupling.  Other possible couplings are forbidden by the conservation of $N_\phi-N_{\phi^\prime}$. This situation can be read off from TABLE \ref{Tab:QuantumNumber2}, where $L$, $N_\phi$, $N_{\phi^\prime}$ and $N_\phi-N_{\phi^\prime}$ for possible Higgs interactions are listed.  However, $N_\phi$ and $N_{\phi^\prime}$ are to be spontaneously broken and a Nambu-Goldstone boson associated with $N_\phi+N_{\phi^\prime}$ is absorbed by the gauge bosons of $SU(2)_L\times U(1)_Y$, but the one associated with $N_\phi-N_{\phi^\prime}$ remains massless. To avoid the appearance of the massless Nambu-Goldstone boson is achieved by introducing a soft breaking term found in TABLE \ref{Tab:QuantumNumber2}.

The use of the conservation of $N_\phi-N_{\phi^\prime}$ organizes mass terms of neutrinos and charged leptons such that $e$-$i$ terms and $i$-$j$ terms ($i,j$=$\mu,\tau$) have different origins, which are translated into the appearance of a generalized $L^\prime$ symmetry, which is identical to $L^\prime$ for $\psi^{e,\mu,\tau}_L$.  It should be noted that another advantage of using $U(1)_{\phi-\phi^\prime}$ lies in the suppression mechanism for a divergent term of $\nu_e\nu_e$ at the two loop level, which would require a tree level mass term of $\nu_e\nu_e$ as a counter term.  This is caused by the interaction of $(h^+h^+)^\dagger \det s$ as depicted in FIG.1(a).  Although this type of the diagram is forbidden by $U(1)_{\phi-\phi^\prime}$, it is in fact allowed by the formation of $\langle 0 \vert \phi^0 \vert 0 \rangle\neq 0$ and $\langle 0 \vert \phi^{\prime 0} \vert 0 \rangle\neq 0$, which yields FIG.1(b) through the interactions of $\phi^{c \dagger}\phi^\prime h^{+\dagger}$ and $\phi^\dagger s\phi^c$.  However, this diagram leads to the finite convergent term.

\section{Neutrino Oscillations}
\subsection{Charged leptons}
Since charged lepton masses include off-diagonal terms, the form of the neutrino mass matrix is affected by the diagonalization process of charged lepton masses so as to maintain diagonal weak currents. The Yukawa couplings of Eq.(\ref{Eq:OurYukawa}) provide the charged lepton mass matrix, $M^\ell_0$, parameterized by
\begin{eqnarray}\label{Eq:OUR_M_ELL}
&
M^\ell_0
=
\left( \begin{array}{ccc}
  0      &  \delta m^\ell_{e\mu}  & \delta m^\ell_{e\tau}\\
  \delta m^\ell_{e\mu}      &  m^0_\mu  & 0\\
  \delta m^\ell_{e\tau}      &  0& m^0_\tau
\end{array} \right),
\end{eqnarray}
arising from $\langle 0 \vert M^\ell_0\left( \phi,\phi^\prime\right)\vert 0 \rangle$ with $M^\ell_0\left( \phi,\phi^\prime\right)$ defined by
\begin{eqnarray}\label{Eq:OUR_M_ELL1}
&
M^\ell_0\left( \phi,\phi^\prime\right)
=
\left( \begin{array}{ccc}
  0      &  f^\phi_\mu \phi  & f^\phi_\tau \phi\\
  f^\phi_\mu \phi      &  f^{\phi^\prime}_\mu\phi^\prime  &0 \\
  f^\phi_\tau \phi      & 0  & f^{\phi^\prime}_\tau\phi^\prime
\end{array} \right),
\end{eqnarray}
where $f^{\phi}_{ie}$ = $f^{\phi}_{ei}$ (=$f^{\phi}_i$) for $i$ = $\mu,\tau$ have been assumed.  The masses of $m$'s and $\delta m$'s are given by
\begin{eqnarray}
&&
\delta m^\ell_{ei} = f^\phi_i v_\phi,
\qquad
m^0_i = f^{\phi^\prime}_iv_{\phi^\prime},
\label{Eq:MassEllEntries}
\end{eqnarray}
for $v_\phi$ ($v_{\phi^\prime}$) = $\langle 0 \vert \phi^0\vert 0 \rangle$ ($\langle 0 \vert \phi^{\prime 0}\vert 0 \rangle$) and should at least satisfy $\vert \delta m^\ell_{ei}\vert\ll \vert m^0_i \vert$ to meet the hierarchy of $m_e\ll m_\mu\ll m_\tau$.  To realize $m_e\ll m_\mu\ll m_\tau$ requires new physics beyond the standard physics; however, we do not intend to discuss how the hierarchy is physically explained but we simply use the parameterization based on the ``hierarchical" one \cite{Hierarchical} to examine its influence on neutrino oscillations \cite{ChargedToNeutral}.

It is straight forward to reach $U_\ell$ that transforms $M^\ell_0$ into $M^\ell=U^\dagger_\ell M^\ell_0 U_\ell$ = diag.($-m_e$, $m_\mu$, $m_\tau$):
\begin{eqnarray}\label{Eq:Unitary}
U_\ell=\left( \begin{array}{ccc}
1-\frac{r^2_{e\mu}+r^2_{e\tau}}{2}  & r_{e\mu} & r_{e\tau}\\
-r_{e\mu} & 1-\frac{r^2_{e\mu}}{2}    & r_{e\mu}r_{e\tau}\frac{m^0_\mu}{m^0_\tau-m^0_\mu} \\
-r_{e\tau}& -r_{e\mu}r_{e\tau}\frac{m^0_\tau}{m^0_\tau-m^0_\mu} & 1-\frac{r^2_{e\tau}}{2}
\end{array} \right),
\end{eqnarray}
up to the second order of $\delta m^\ell$'s contained in $r_{e\mu,e\tau}$: 
\begin{eqnarray}\label{Eq:UnitaryElements}
&&
r_{e\mu}=
\frac{\delta m^\ell_{e\mu}}{m^0_\mu},
\quad
r_{e\tau}=\frac{\delta m^\ell_{e\tau}}{m^0_\tau}
\end{eqnarray}
and the diagonal masses are calculated to be
\begin{eqnarray}
&&
m_e = m^0_\mu r^2_{e\mu}+ m^0_\tau r^2_{e\tau},
\quad 
m_\mu = m^0_\mu \left( 1+ r^2_{e\mu}\right),
\quad 
m_\tau = m^0_\tau \left( 1+ r^2_{e\tau}\right),
\label{Eq:DiagonalChargedLeptons}
\end{eqnarray}
which give the upper bounds on $r_{e\mu,e\tau}$:
\begin{eqnarray}
&&
\vert r_{e\mu}\vert \mapleq \sqrt{m_e/m_\mu},
\quad
\vert r_{e\tau}\vert \mapleq \sqrt{m_e/m_\tau}, 
\label{Eq:DiagonalChargedLeptons2}
\end{eqnarray}
respectively, from $m_e \geq r^2_{e\mu}m^0_\mu\approx r^2_{e\mu}m_\mu$ and $m_e \geq r^2_{e\tau}m^0_\tau\approx r^2_{e\tau}m_\tau$.  

Even after the rotation of the mass term given by $U^\dagger_\ell \langle 0 \vert M^\ell_0 ( \phi,\phi^\prime )\vert 0 \rangle U_\ell$, our Yukawa interactions corresponding to $U^\dagger_\ell M^\ell_0 ( \phi,\phi^\prime )U_\ell$ $(=M^\ell ( \phi,\phi^\prime ))$ contain flavor-off-diagonal couplings.  We find that $M^\ell\left( \phi,\phi^\prime\right)$ is calculated to be:
\begin{eqnarray}\label{Eq:OUR_YUKAWA}
M^\ell\left( \phi,\phi^\prime\right)
&=&
U^\dagger_\ell M^\ell_0 \left( \phi,\phi^\prime \right)U_\ell
\nonumber \\
&=&
\left( \begin{array}{cc}
-\left( 2\alpha_\phi-\alpha_{\phi^\prime}\right)m_e
&
\left( \alpha_\phi-\alpha_{\phi^\prime}\right)r_{e\mu}m^0_\mu
\\
\left( \alpha_\phi-\alpha_{\phi^\prime}\right)r_{e\mu}m^0_\mu
&
\left[\alpha_{\phi^\prime}+\left( 2\alpha_\phi-\alpha_{\phi^\prime}\right)r^2_{e\mu}\right]m^0_\mu
\\
\left( \alpha_\phi-\alpha_{\phi^\prime}\right)r_{e\tau}m^0_\tau
&
\left( \alpha_\phi-\alpha_{\phi^\prime}\right)r_{e\mu}r_{e\tau}\left(m^0_\mu+m^0_\tau\right)
\end{array} \right.
\nonumber \\
&& \hspace{30mm}\left. \begin{array}{c}
\left( \alpha_\phi-\alpha_{\phi^\prime}\right)r_{e\tau}m^0_\tau
\\
\left( \alpha_\phi-\alpha_{\phi^\prime}\right)r_{e\mu}r_{e\tau}\left(m^0_\mu+m^0_\tau\right)
\\
\left[\alpha_{\phi^\prime}+\left( 2\alpha_\phi-\alpha_{\phi^\prime}\right)r^2_{e\tau}\right]m^0_\tau
\end{array} \right),
\end{eqnarray}
where $\alpha_\phi=\phi/v_\phi$ and $\alpha_{\phi^\prime}=\phi^\prime/v_{\phi^\prime}$, which induces flavor-changing interactions for $\tau$ and $\mu$.  Of course, Eq.(\ref{Eq:OUR_YUKAWA}) with the identification of $\alpha_{\phi^\prime}$ with $\alpha_\phi$, corresponding to the case of the standard model, only contains diagonal Higgs couplings giving rise to diag.($-m_e$, $m_\mu$, $m_\tau$). The flavor-changing interactions are roughly controlled by the suppression factor of order $r_{e\mu}r_{e\tau}(m_\mu/v_{weak})(m_\tau/v_{weak})(v_{weak}/m)^2$ (=$\xi_1$) for $\tau \to \mu\mu\mu$, $\mu ee$, $\mu\gamma$, $r^2_{e\mu}r_{e\tau}(m_\mu/v_{weak})(m_\tau/v_{weak})(v_{weak}/m)^2$ (=$\xi_2$) for $\tau \to \mu\mu e$, $e\gamma$ and $r_{e\mu}(m_e/v_{weak})(m_\mu/v_{weak})(v_{weak}/m)^2$ (=$\xi_3$) for $\mu \to eee$, $e\gamma$, where $m$ is a mass of the mediating Higgs scalar and $v_\phi\sim v_{\phi^\prime} \sim v_{weak}$ = $( 2{\sqrt 2}G_F)^{-1/2}$=174 GeV. We find that to suppress these interactions to the phenomenologically consistent level requires a rough estimate of $\vert\xi_{1,2,3}\vert \mapleq 10^{-5}$ \cite{NewWork}, which can be fulfilled because of $m \mapgeq v_{weak}$ and Eq.(\ref{Eq:DiagonalChargedLeptons2}) for $r_{e\mu,e\tau}$.  It should be noted that there are no such Higgs interactions for quarks that only couple to $\phi$. The similar flavor-changing interactions caused by $h^+$ \cite{h_phenom2} are sufficiently suppressed because of the smallness of the $h^+$-couplings to leptons to be estimated in Eq.(\ref{Eq:CouplingValues}).

\subsection{Neutrinos}
To estimate effects on the neutrino mass matrix through the rotation due to $U_\ell$, we shift the original base into the one with the diagonalized charged lepton masses, which forces us to rotate the original neutrinos ($\vert\nu\rangle$) into $\vert\nu_{weak}\rangle$ in the weak base by $\vert\nu_{weak}\rangle = U^\dagger_\ell \vert\nu\rangle$. The radiative neutrino masses to be denoted by  $\delta m^\nu_{ij}$ are generated by interactions corresponding to FIG.2.  For the sake of simplicity, we set $v_\phi$ = $v_{\phi^\prime}$ and $m_\phi$ = $m_{\phi^\prime}$,
\footnote{Our later discussions are still valid as far as $\vert rK_\phi - r^{-1}K_{\phi^\prime}\vert \mapleq\vert K_{\phi^\prime}\vert/10$ for $r = v_{\phi^\prime}/v_\phi$. For instance, $a$ (=$\delta m^\nu_{ee}$) is computed to be $2(rK_\phi-r^{-1}K_{\phi^\prime})(r_{e\mu}-\sigma r_{e\tau})f_{[e\mu]}m^2_\mu$ for $f_{[e\mu]}m^2_\mu = -\sigma f_{[e\tau]}m^2_\tau$, which is at most $r_{e\mu}b/10$ being harmless for our estimation, and similarly for other elements.} 
 where $m_{\phi , \phi^\prime }$ denote the masses of $\phi^+$ and $\phi^{\prime +}$, and calculate $\delta m^\nu_{ij}$ to be:
\begin{eqnarray}
\delta m^\nu_{ij} 
&=& 
K_{\phi^\prime}\left( U^T_\ell {\bf f}  M^\ell_0 M^{\ell\dagger}_0 U_\ell\right)_{ij}-(i\leftrightarrow j),
\label{Eq:RadMassNu}
\end{eqnarray}
for $\vert\nu_{weak}\rangle$, where ${\bf f}_{ij} = f_{[ij]}$ and $K_{\phi^\prime}$ is the one-loop factor:
\begin{eqnarray}
&&
K_{\phi^\prime} = \frac{\mu_0}{16\pi^2}
    \frac{\ln m^2_h-\ln m^2_{\phi^\prime}}{m^2_h-m^2_{\phi^\prime}},
\label{Eq:KineticFactor}
\end{eqnarray}
where $m_h$ is a mass of the $h^+$ scalar. We find that
\begin{eqnarray}
&&
\delta m^\nu_{ij} 
= 
K_{\phi^\prime}F_{[ij]}\left( m^2_{\ell^j}-m^2_{\ell^i}\right),
\label{Eq:RadMassNuEntries}
\end{eqnarray}
where the couplings of $F_{[ij]}$ are defined by $F_{[ij]}$=$(U^T_\ell {\bf f}U_\ell)_{ij}$, which result in 
\begin{eqnarray}
&&
F_{[e\mu]} = f_{[e\mu]},
\quad
F_{[e\tau]} = f_{[e\tau]},
\quad
F_{[\mu\tau]} = r_{e\mu}f_{[e\tau]}-r_{e\tau}f_{[e\mu]}.
\label{Eq:RotatedCoupling1}
\end{eqnarray}
The tree level masses, $m^\nu_{ij}$, are given by the type II seesaw mechanism to be: 
\begin{eqnarray}\label{Eq:TreeMassNuEntries} 
&&
m^\nu_{ij} =f_{\{ij\}} v_s \approx f_{\{ij\}}\mu \frac{v^2_\phi}{2m^2_s},
\label{Eq:TreeLevelMass}
\end{eqnarray}
where $f_{\{ \mu\mu\}} = f_{\{ \tau\tau\}} = \sigma f_{\{ \mu\tau\}}$ ($\equiv f_s$) and $v_s$ = $\langle 0 \vert s^0 \vert 0 \rangle$.

Collecting these results, we find that our mass matrix of Eq.(\ref{Eq:M_NU}) has the following mass parameters:
\begin{eqnarray}\label{Eq:MassMatrixElement}
&&
a = 0,
\quad
b = \delta m^\nu_{e\mu} = K_{\phi^\prime} f_{[e\mu]}m^2_\mu,
\quad
c = \delta m^\nu_{e\tau} = K_{\phi^\prime} f_{[e\tau]}m^2_\tau,
\nonumber \\
&&
d = m^\nu,
\quad
e = \sigma m^\nu + \delta m^\nu_{\mu\tau} = \sigma m^\nu+K_{\phi^\prime} F_{[\mu\tau]}m^2_\tau,
\quad
f = m^\nu,
\end{eqnarray}
where $m^\nu = f_sv_s$ and $m^2_e \ll m^2_{\mu,\tau}$ has been used to estimate Eq.(\ref{Eq:RadMassNuEntries}). The two-loop convergent contribution shown in FIG.1(b) to the $\nu_e\nu_e$ mass for the $a$-term is well suppressed by the presence of $m_s$ contained in the propagator of $s$ and does not jeopardize $a$=0.  The maximal atmospheric neutrino mixing defined by $\vert t_{23}\vert =1$ calls for the ``inverse" hierarchy of $f_{[e\mu]}$ and $f_{[e\tau]}$ \cite{InverseCoupling} expressed as 
\begin{eqnarray}\label{Eq:Requirement1}
&
f_{[e\mu]}m^2_\mu = -\sigma f_{[e\tau]}m^2_\tau.
\end{eqnarray}
  For the solar neutrino oscillations, the radiative $\nu_\mu$-$\nu_\tau$ mixing term of $F_{[\mu\tau]}$ arises via $r_{e\mu,e\tau}$ as a result of the rotation due to the $e$-$\mu$ and $e$-$\tau$ mixings.  It is this term that gives a deviation of $\sin^22\theta_{12}$ from unity.  The masses of neutrinos satisfy the ``normal" mass hierarchy of $\vert m_1\vert < \vert m_2\vert \ll m_3$ determined by Eq.(\ref{Eq:NuMasses123}) to be:
\begin{eqnarray}\label{Eq:EstimatedNuMasses}
&&
m_1 = -\eta\frac{ \sqrt{8+x^2}-\vert x\vert}{2}\delta m^\nu_{rad},
\quad
m_1 = \eta\frac{ \sqrt{8+x^2}+\vert x\vert}{2}\delta m^\nu_{rad},
\nonumber \\
&&
m_3 = 2 m^\nu+\eta x\delta m^\nu_{rad}
\end{eqnarray}
with $\delta m^\nu_{rad}$ = $\sqrt{(\delta m^{\nu 2}_{e\mu}+\delta m^{\nu 2}_{e\tau})/2}$, where the sign of $\eta$ has been adjusted to yield $\eta x =-\vert x \vert$ for $\vert m_1 \vert <  \vert m_2 \vert$ and $x$ measures the ratio of the $\nu_\mu$-$\nu_\tau$ mixing mass over the $\nu_e$-$\nu_\tau$ mixing mass, which is defined by 
\begin{eqnarray}\label{Eq:Ratio_Of_x}
&&
x =\frac{\delta m^\nu_{\mu\tau}}{\vert\delta m^\nu_{e\tau}\vert} = \frac{F_{[\mu\tau]}}{\vert F_{[e\tau]}\vert} = \frac{r_{e\mu}F_{[e\tau]}-r_{e\tau}F_{[e\mu]}}{\vert F_{[e\tau]}\vert}.
\end{eqnarray}
Then, $\Delta m^2_{atm,\odot}$ and $\sin^22\theta_{12}$ are calculated to be:
\begin{eqnarray}\label{Eq:Sin_12}
&&
\Delta m^2_{atm} \approx 4m^{\nu 2}+4\eta xm^\nu\delta m^\nu_{rad},
\quad
\Delta m^2_\odot \approx \vert x\vert \sqrt{8+x^2}\delta m^{\nu 2}_{rad},
\nonumber \\
&&
\sin^22\theta_{12} = \frac{8}{\left( 8+x^2\right)}.
\end{eqnarray}

To get an estimate of $x$, we parameterize $r_{e\mu,e\tau}$ as 
\begin{eqnarray}\label{Eq:ElectronMass}
&&
r_{e\mu} = c_\ell \sqrt{\frac{m_e}{m_\mu}},
\quad
r_{e\tau} = s_\ell \sqrt{\frac{m_e}{m_\tau}}
\end{eqnarray}
to satisfy the constraint of Eq.(\ref{Eq:DiagonalChargedLeptons}) on the electron mass, where $c_\ell = \cos\theta_\ell$ and $s_\ell = \sin\theta_\ell$.  The parameter of $x$ turns out to be given by
\begin{eqnarray}\label{Eq:Ratio_Of_x2}
&
\vert x \vert= \vert 0.07c_\ell +4.86\sigma s_\ell\vert, 
\end{eqnarray}
where the hierarchical coupling condition of Eq.(\ref{Eq:Requirement1}) has been used.  For $\vert x\vert$ = $\sqrt{2}$ corresponding to $s_\ell \sim \pm 0.3$, $\sin^22\theta_{12}$ = 0.8 and $\delta m^\nu_{rad}$ = 3.2$\times 10^{-3}$ eV are obtained.  The tree level mass of $m^\nu$ is estimated to be $\sim 0.027$ eV for $\Delta m^2_{atm} = 3 \times 10^{-3}$ eV$^2$.  The type II seesaw mechanism yields an estimate of the mass parameter: $m_s$ ($= \mu$) = 1.2$\times 10^{14} \times (\vert f_{\{ij\}}\vert /e)$ GeV for $v_\phi$ (= $v_{\phi^\prime}$)=$v_{weak}/\sqrt{2}$ to meet $v^2_\phi + v^2_{\phi^\prime} = v^2_{weak}$ for the weak boson masses, where $e$ is the electromagnetic coupling.  From Eq.(\ref{Eq:Requirement1}) for the maximal atmospheric neutrino mixing, 
\begin{eqnarray}\label{Eq:CouplingValues}
&&
f_{[e\mu]}\sim 2.9\times 10^{-5},
\quad
f_{[e\tau]}\sim - \sigma\times 10^{-7},
\end{eqnarray}
together with $\vert F_{[\mu\tau]}\vert= \sqrt{2}\vert f_{[e\tau]}\vert$ are obtained, where $\mu_0$ = $m_\phi$ = $m_{\phi^\prime}$ = $v_{weak}$ and $m_{h^+}$ = $3v_{weak}$ are used to compute the loop-factor of $K_{\phi^\prime}$.  The masses of $m_{1,2,3}$ are predicted to be: 
\begin{eqnarray}\label{Eq:MassiveMass}
&&
\vert m_1 \vert = 2.8 \times 10^{-3}~{\rm eV},
\quad 
\vert m_2 \vert = 7.3 \times 10^{-3}~{\rm eV},
\quad
m_3 = 5.5 \times 10^{-2}~{\rm eV}.
\end{eqnarray}

\section{Summary}
Summarizing our discussions, we have demonstrated that the use of $U(1)_\phi$ and $U(1)_{\phi^\prime}$ successfully opens a window for both $\sin^22\theta_{23}=1$ for atmospheric neutrino oscillations and the LMA solution with $\sin^22\theta_{12} \sim 0.8$ for solar neutrino oscillations.  The mass mixings in the $e$-$\mu$ and $e$-$\tau$ entries in the neutrino mass matrix, {\it i.e.} the $\nu_e$-$\nu_\mu$ and $\nu_e$-$\nu_\tau$ mixing masses of $\delta m^\nu_{e\mu}$ and $\delta m^\nu_{e\tau}$, determine the atmospheric neutrino mixing: $\tan \theta_{23}$ = $-\delta m^\nu_{e\tau}/\delta m^\nu_{e\mu}$ and supply solar neutrino masses proportional to $\delta m^\nu_{rad}$ (=$\sqrt{(\delta m^{\nu 2}_{e\mu}+\delta m^{\nu 2}_{e\tau})/2}$). On the other hand, mass mixings in the charged lepton mass matrix, {\it i,e.} the $e$-$\mu$ and $e$-$\tau$ mixing masses of $\delta m^\ell_{e\mu}$ and $\delta m^\ell_{e\tau}$, are rotated into the electron mass given by $\delta m^{\ell 2}_{e\mu}/m_\mu$+$\delta m^{\ell 2}_{e\tau}/m_\tau$ and induce the $\delta m^\nu_{\mu\tau}$-mixing mass as effects of the rotation, which yields the deviation of $\sin^22\theta_{12}$ from unity. This induced $\nu_\mu$-$\nu_\tau$ mixing determines $\sin^22\theta_{12} = 8/(8+x^2)$ for $x=\delta m^\nu_{\mu\tau}/\vert\delta m^\nu_{e\tau}\vert$, leading to $\sin^22\theta_{12} = 0.8$ for $\vert x\vert$ = $\sqrt{2}$, which corresponds to the mixing angle of $s_\ell$ $\sim$ $\pm$0.3 that measures the $\tau$-contribution in the electron mass as in Eq.(\ref{Eq:ElectronMass}).  To induce the observed neutrino oscillations is thus based on the radiative mechanism and the rotation process to get the electron mass, which can be illustrated by
\begin{eqnarray}
\left( \begin{array}{ccc}
  0            &  0            & 0\\
  0            &  m^\nu            & \sigma m^\nu\\
  0            &  \sigma m^\nu     & m^\nu
\end{array} \right) 
&\stackrel{\mathrm{radiative}}{\Longrightarrow}&
\left( \begin{array}{ccc}
  0            &  \delta m^\nu_{e\mu}   & \delta m^\nu_{e\tau}\\
  \delta m^\nu_{e\mu}            &  m^\nu            & \sigma m^\nu\\
  \delta m^\nu_{e\tau}           &  \sigma m^\nu     & m^\nu
\end{array} \right) 
\nonumber \\
&\stackrel{\mathrm{rotation}}{\Longrightarrow}&
\left( \begin{array}{ccc}
  0            &  \delta m^\nu_{e\mu}   & \delta m^\nu_{e\tau}\\
  \delta m^\nu_{e\mu}            &  m^\nu            & \sigma m^\nu+\delta m^\nu_{\tau\mu}\\
  \delta m^\nu_{e\tau}     &  \sigma m^\nu+\delta m^\nu_{\tau\mu}     & m^\nu
\end{array} \right).
\nonumber
\end{eqnarray}
Once the democratic mass structure for $\nu_\mu$ and $\nu_\tau$ as the $m^\nu$-terms is realized, we can generate radiative neutrino masses whose texture is compatible with the large solar neutrino mixing.

\begin{center}
{\bf Acknowledgements}
\end{center}

The authors are grateful to Y. Koide for valuable discussions.  One of the authors (M.Y.) also thanks to the organizers and participants in Summer Institute 2001 at FujiYoshida, Yamanashi, Japan, for useful comments.  The work of M.Y. is supported by the Grants-in-Aid for Scientific Research on Priority Areas A, "Neutrino Oscillations and Their Origin," (No 12047223) from the Ministry of Education, Culture, Sports, Science, and Technology, Japan.


\vspace{4mm}
\centerline{TABLES}
\begin{table}[htbp]
    \caption{\label{Tab:QuantumNumber1}The lepton number ($L$), $N_\phi,N_{\phi^\prime}$ and $N_\phi\pm N_{\phi^\prime}$ for leptons and Higgs scalars, where $N_\phi+N_{\phi^\prime}$ is nothing but the hypercharge.}
    \begin{center}
    \begin{tabular}{ccccccccc}\hline
                           & $\psi^e_L$ & $\psi^\mu_L,\psi^\tau_L$ & $e_R$ & $\mu_R,\tau_R$ & $\phi$ & $\phi^\prime$ & $h^+$ & $s$\\ \hline
  $L$                      & 1    &  1   &  1   & 1    & 0 & 0    & $-2$ & $-2$\\
  $N_\phi$                 & 0    & $-1$ & $-2$ & $-1$ & 1 & 0    & 1    & 2\\
  $N_{\phi^\prime}$        & $-1$ & 0    & 0    & $-1$ & 0 & 1    & 1    & 0\\
  $N_\phi+N_{\phi^\prime}$ & $-1$ & $-1$ & $-2$ & $-2$ & 1 & 1    & 2    & 2\\
  $N_\phi-N_{\phi^\prime}$ & 1    & $-1$ & $-2$ & 0    & 1 & $-1$ & 0    & 2\\ \hline
    \end{tabular}
    \end{center}
\end{table}

\begin{table}[htbp]
    \caption{\label{Tab:QuantumNumber2}$L$, $N_\phi,N_{\phi^\prime}$ and $N_\phi-N_{\phi^\prime}$ for Higgs interactions, where $N_\phi+N_{\phi^\prime}$ = 0.}
    \begin{center}
    \begin{tabular}{cccccc}\hline
      & $\phi^{c \dagger}\phi^\prime h^{+\dagger}$ 
      & $\phi^{\prime\dagger}\phi$ 
      & $\left( h^+h^+\right)^\dagger \det s$ 
      & $\phi^{c \dagger}s\phi^c h^{+\dagger}$ 
      & $\phi^{c \dagger}s\phi^{\prime c}h^{+\dagger}$ 
      \\ \hline
$L$
      & 2 
      & 0
      & 0
      & 0
      & 0
      \\
$N_\phi$
      & 0 
      & 1
      & 2
      & 1
      & 0
      \\
$N_{\phi^\prime}$
      & 0 
      & $-1$
      & $-2$
      & $-1$
      & 0
      \\
$N_\phi-N_{\phi^\prime}$
      & 0 
      & 2
      & 4
      & 2
      & 0
      \\ \hline
      & $\phi^{\prime c\dagger}s\phi^c h^{+\dagger}$ 
      & $\phi^{\prime c\dagger}s\phi^{\prime c}h^{+\dagger}$
      & $\phi^\dagger s\phi^c$ 
      & $\phi^\dagger s\phi^{\prime c}$ 
      & $\phi^{\prime \dagger}s\phi^{\prime c}$
      \\ \hline
$L$
      & 0
      & 0
      & $-2$
      & $-2$
      & $-2$
      \\
$N_\phi$
      & 2
      & $-1$
      & 0
      & 1
      & 2
      \\
$N_{\phi^\prime}$
      & $-2$
      & 1
      & 0
      & $-1$
      & $-2$
      \\
$N_\phi-N_{\phi^\prime}$
      & 4
      & $-2$
      & 0
      & 2
      & $-4$
      \\ \hline
    \end{tabular}
    \end{center}
\end{table}

\noindent
\begin{figure}[htbp] 
\vspace*{13pt}
\centerline{\epsfbox{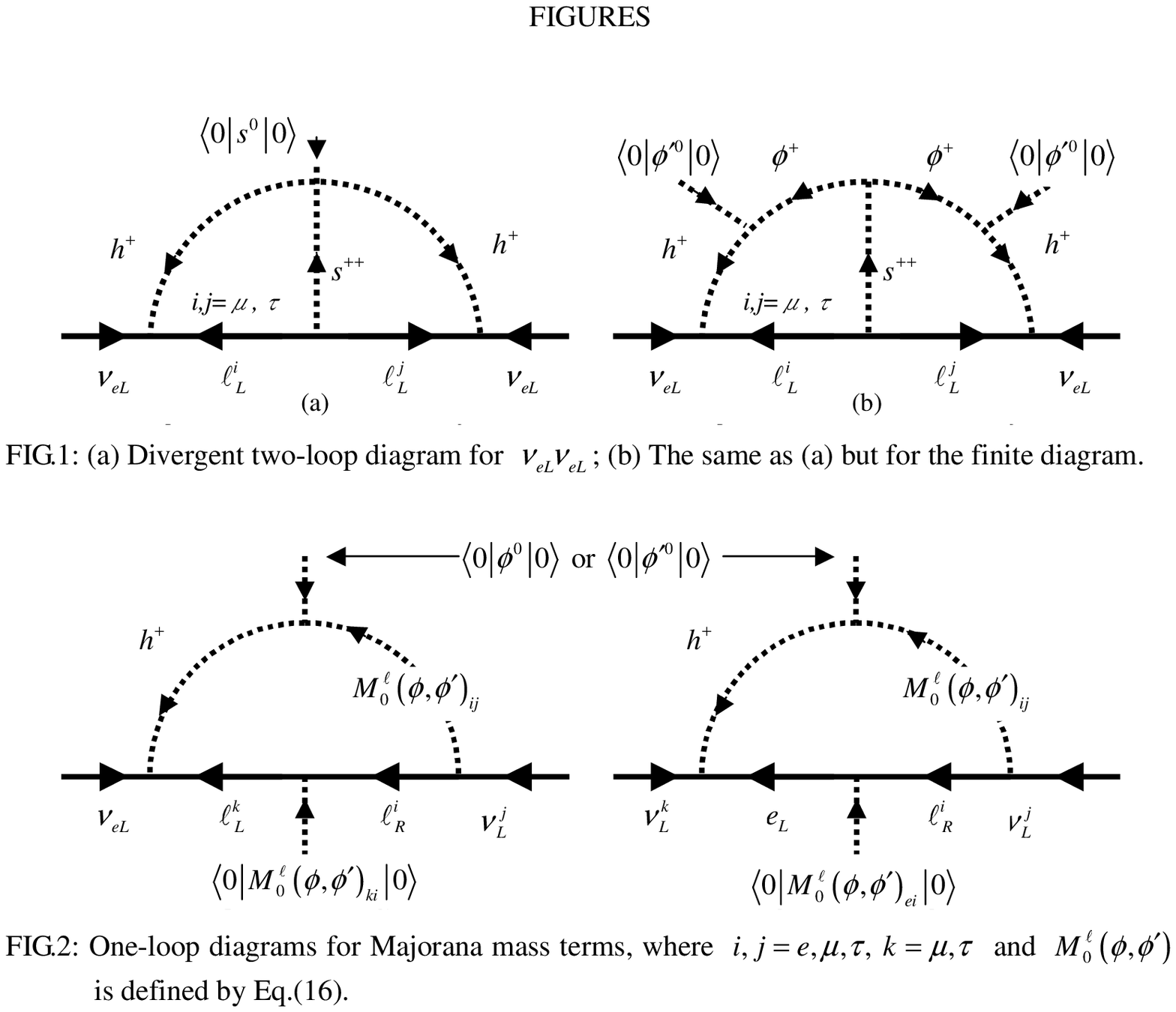}}
\end{figure}

\end{document}